# Scientific evaluation of Charles Dickens

Mikhail Simkin

**Abstract:** I report the results of the test, where the takers had to tell the prose of Charles Dickens from that of Edward Bulwer-Lytton, who is considered by many to be the worst writer in history of letters. The average score is about 50%, which is on the level of random guessing. This suggests that the quality of Dickens' prose is the same as of that of Bulwer-Lytton.

F. Scott Fitzgerald wrote, "Let me tell you about the very rich. They are different from you and me." On this Ernest Hemingway commented, "Yes, they have more money."

Are the very famous writers different from the obscure ones?

The question may seem shocking to some people, but recent scientific research makes it quite reasonable. The study of misprints in scientific citations had lead to the conclusion that about 80% of citations are copied from the lists of references used in other papers [1]. Thus, in a majority of cases, a citation is not a result of an independent evaluation of the qualities of the cited paper but merely an imitation of another citer's behavior. This way a paper that already was cited is likely to be cited again, and after it is cited again it is even more likely to be cited in the future. Thus some papers can become much more cited than others even when identical in merit. Mathematical modeling of the process of citation copying demonstrated that major features of the citation distribution could be explained even under assumption that all papers are created equal [2]. One can suspect that, similar to highly cited scientists, highly popular writers can become such as a result of the ordinary law of chances. One way to check that is to see if people can appreciate the prose of a famous writer when his name is detached from it.

Edward Bulwer-Lytton is the worst writer in history of letters. An annual wretched writing contest [3] was established in his honor. In contrast, Charles Dickens is one of the best writers ever. Can one tell the difference between their prose? To check this I wrote the "Great prose or not?" quiz [4]. It consists of a dozen of representative literary passages, written either by Bulwer-Lytton or by Dickens. The takers are to choose the author of each quote. They face a formidable task and are often surprised to learn the correct answers. "What mindless boob would write such tripe? Dickens, one would know now." – wrote me one of respondents. The distribution of the scores received by over nine thousands quiz-takers[1] is shown in Figure 1. The average score is 5.78 or 48.2% correct.

---

[1] When I looked at the quiz results I noticed hundreds of cases when two or more scores came from the same IP address within few minutes. In many of such cases the later score was 100%. This suggests that many people took several shots at the quiz. To eliminate this cheating I cleaned the data by selecting only the first score from each IP address. Afterwards I cleaned the data from the results, where one or more questions were skipped.

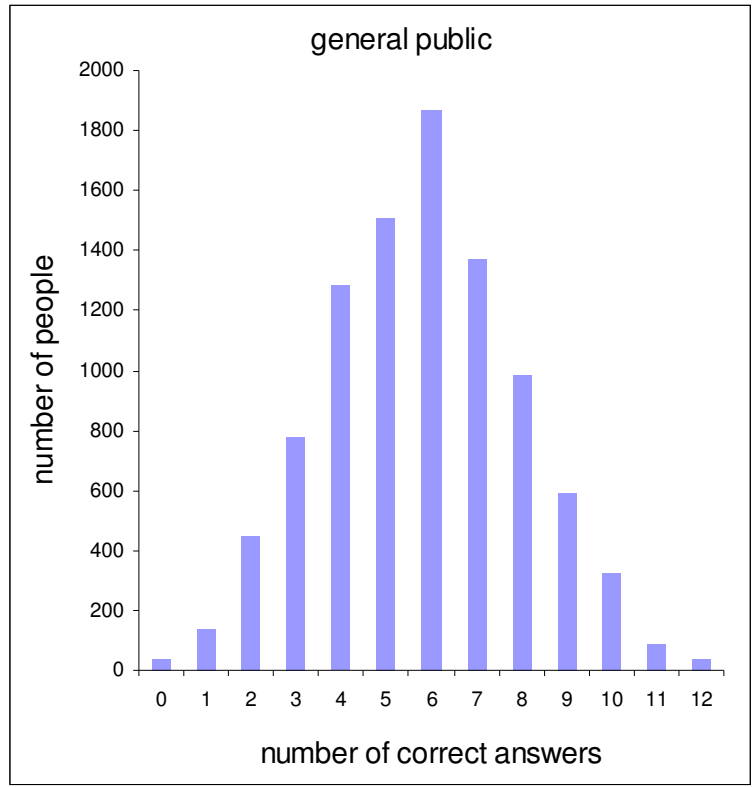

**Figure 1.** The histogram of the scores earned by 9,461 people on the "Great prose or not?" quiz. The average score is 5.78 or 48% correct. The standard error of this average is 0.022 or 0.19%.

**Table 1.** Fraction of people who attributed each quote to Dickens and to Bulwer-Lytton, along with the true author.

| Question number | The real author, and the book the excerpt is taken from | Selected as Dickens | Selected as Bulwer-Lytton |
|---|---|---|---|
| 1 | Charles Dickens, "Great Expectations" | 42.5% | 57.5% |
| 2 | Edward Bulwer-Lytton, "Eugene Aram" | 50.5% | 49.5% |
| 3 | Charles Dickens, "Great Expectations" | 54.6% | 45.4% |
| 4 | Edward Bulwer-Lytton, "Eugene Aram" | 49.9% | 50.1% |
| 5 | Charles Dickens, "David Copperfield" | 50.7% | 49.3% |
| 6 | Edward Bulwer-Lytton, "Eugene Aram" | 50.1% | 49.9% |
| 7 | Charles Dickens, "Great Expectations" | 59.9% | 40.1% |
| 8 | Edward Bulwer-Lytton, "Paul Clifford" | 49.6% | 50.4% |
| 9 | Edward Bulwer-Lytton, "Eugene Aram" | 36.5% | 63.5% |
| 10 | Charles Dickens, "Great Expectations" | 40.6% | 59.4% |
| 11 | Charles Dickens, "David Copperfield" | 40.8% | 59.2% |
| 12 | Edward Bulwer-Lytton, "Paul Clifford" | 74.3% | 25.7% |

There are two possible answers to each test question. If one is completely clueless and resorts to random guessing, he will on average get 50% of the questions right. With the average score of 48% our quiz-takers lost to a monkey. On average, a quote from Bulwer-Lytton was selected as Dickens (or great prose) by 52% of quiz-takers, while a quote from Dickens was selected as Dickens by only 48%. Does this mean that Bulwer-Lytton is a better writer than Dickens? Probably not. Table 1 shows for every quote the fraction of people who attributed it to Dickens. This fraction varies between the quotes with the lowest being 36% (No. 9) and the highest 74% (No. 12). This suggests that a different selection of quotes could lead to a different average score. For example, if we remove the most Dickensian Bulwer (No. 12) and the most Bulwerian Dickens (No. 10), and recalculate the scores based on 10 remaining questions, - the average score becomes 51%.

An interesting thing is that out of 9,461 people, 38 got every question wrong and 37 got everything right. The approximate equality of these numbers is consistent with random guessing, but their magnitude is not. It is more than fifteen times bigger than what random guessing would give. The explanation is that some of quiz takers can sniff stylistic similarities between certain literary passages and attribute them to the same writer. This would help them to get a higher score, if they can determine which writer is good and which is bad, otherwise they are equally likely to get a very high or a very low score.

The performance of our quiz-takers is bad. But could this be because they don't know English? The feedback demonstrates, however, that even educated people can't tell Dickens from Bulwer. One of quiz-takers wrote me "I got a 50%. My cat could do that well. The wine experts say a peek at the label is worth a thousand sips, and that seems to hold here. As a classicist I'm frequently called on to teach stuff I think is wretched, just because it's old." Some experts do not even dare to take the quiz. Prof. Scott Rice, the founder of the Bulwer-Lytton fiction contest [3], wrote to me "I haven't really taken it yet myself. Perhaps I am afraid to." We can address the education issue in a scientific way. Fortunately, the quizzing script records taker's IP address. From it, one can infer where their computers were located. I selected a subset of scores, which were received by people coming from English-speaking (American, British, Australian, and New Zealandian) universities[2]. The histogram of the scores received by 602 such people is shown in Figure 2. The average score is 5.76 or 48.0% correct. The standard error of this average is 0.095 or 0.8%. Educated English-speaking folks lost to the general public, whose average score is 48.2%. The difference between the scores is, however, statistically insignificant, because it is less than the standard error.

But, perhaps, just knowing English is not enough? May be the beauty of Dickens' prose is so far beyond the apprehensions of the vulgar that only the most cultured people can appreciate it? To check this I selected a subset of scores, earned by people coming from elite universities (Ivy League and Oxbridge). Table 2 contains the statistics of scores received by 76 of the chosen. The average score is 6 or 50% correct. The elite won over crowd by the whole 2%. The difference between the elite and general scores is, however, statistically insignificant. Due to the small size of the elite sample the standard error of the average elite score is 2.6%.

---

[2] I identified American, British, Australian, and New Zealandian universities by Internet domains: edu, ac.uk, edu.au, and ac.nz.

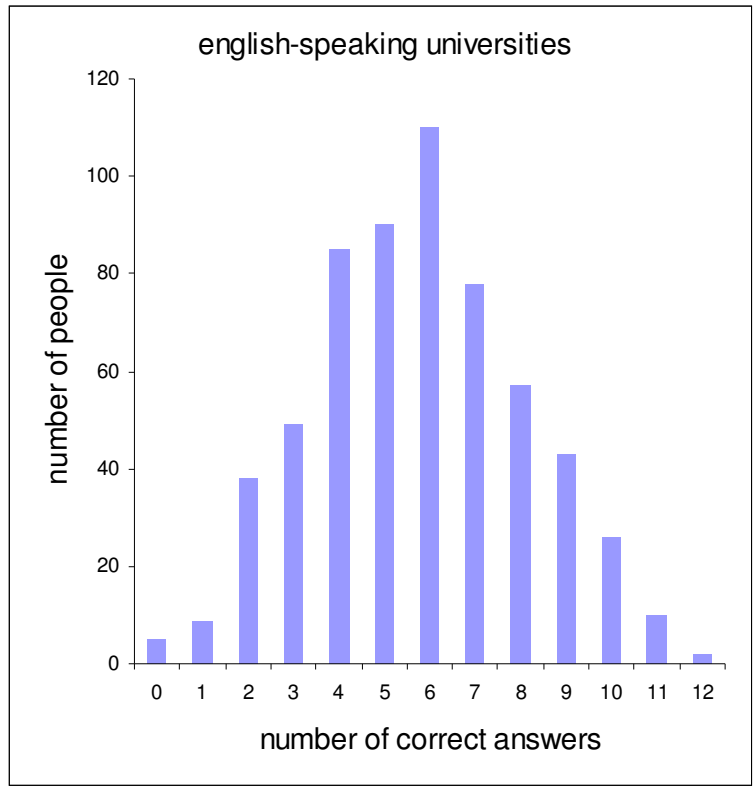

**Figure 2.** The histogram of the test scores earned by 602 people, coming from American, British, Australian, and New Zealandian universities. The average score is 5.76 or 48.0% correct. The standard error of this average is 0.095 or 0.8%.

**Table 2.** Statistics of the elite (Ivy League and Oxbridge) scores on "Dickens or Bulwer-Lytton?" quiz. The average elite score is 6 or 50% correct. The standard error of this average is 0.3 or 2.6%.

| Elite School | number of respondents | minimum score | maximum score | average score |
|---|---|---|---|---|
| Brown University | 2 | 3 | 6 | 4.50 |
| Columbia University | 13 | 2 | 9 | 5.08 |
| Cornell University | 3 | 4 | 7 | 6.00 |
| Harvard University | 14 | 1 | 10 | 5.71 |
| Princeton University | 2 | 3 | 9 | 6.00 |
| University of Cambridge | 16 | 2 | 9 | 6.13 |
| University of Oxford | 10 | 3 | 10 | 6.30 |
| University of Pennsylvania | 7 | 2 | 11 | 7.71 |
| Yale University | 9 | 3 | 11 | 6.56 |
| Total | 76 | 1 | 11 | 6.00 |

Of course, the method used in this investigation has limitations. A novel is characterized not just by its prose style, but also by its plot and characters. The method compares only prose styles. Note, however, that the Bulwer-Lytton quotation, used as an epigraph for the wretched writing contest [3] is just as long as those used in the quiz. At the very least results of the quiz show that the contest could well be named after Dickens. In addition, in a related experiment, publishers had rejected Booker prize-winning novels submitted as works by aspiring authors [5]. The publishers were given the whole chapters of the books, not just paragraphs, but still failed to spot great prose.

I began this paper with the question: Are the famous writers different from their obscure colleagues? The answer is: Yes, they have more readers.

I reported similar results for the case of Modern Art [6].